\documentclass {jaa}
\usepackage[english]{babel}
\usepackage{times,amssymb,subfigure,relsize}
\usepackage{color,rotating}
\usepackage{amssymb}
\usepackage{comment}
\usepackage{graphicx}
\usepackage{hyperref}
\usepackage{graphics}
\usepackage{graphicx}
\usepackage{color}
\usepackage{amsmath}
\usepackage{psfrag}
\usepackage{epsfig}
\usepackage{multirow}
\usepackage{authblk}

\begin{document}
\title{Large Area X-ray Proportional Counter (LAXPC) Instrument on {\it AstroSat} and Some Preliminary Results from its performance in the orbit}
\author{P. C. Agrawal\textsuperscript{1}, J. S. Yadav\textsuperscript{2,*}, H. M. Antia\textsuperscript{2}, Dhiraj Dedhia\textsuperscript{2}, P. Shah\textsuperscript{2}, Jai Verdhan Chauhan\textsuperscript{2}, R. K. Manchanda\textsuperscript{3},  V. R. Chitnis\textsuperscript{2},  V. M. Gujar\textsuperscript{2}, Tilak Katoch\textsuperscript{2}, V. N. Kurhade\textsuperscript{2}, 
P. Madhwani\textsuperscript{2}, T. K. Manojkumar\textsuperscript{2}, V. A. Nikam\textsuperscript{2}, A. S. Pandya \textsuperscript{2}, J. V. Parmar\textsuperscript{2}, D. M. Pawar\textsuperscript{2}, Jayashree Roy\textsuperscript{1}, B. Paul\textsuperscript{4}, Mayukh Pahari\textsuperscript{5}, Ranjeev Misra\textsuperscript{5}, M. H. Ravichandran\textsuperscript{6}, K. Anilkumar\textsuperscript{6}, C. C. Joseph\textsuperscript{6}, K. H. Navalgund\textsuperscript{7}, R. Pandiyan\textsuperscript{7}, K. S. Sarma\textsuperscript{7},  K. Subbarao\textsuperscript{7}}
\affilOne{\textsuperscript{1}UM-DAE Center of Excellence for Basic Sciences, University of Mumbai Campus at Vidhyanagar, Kalina, Mumbai-400098, India\\}
\affilTwo{\textsuperscript{2}Tata Institute of Fundamental Research, Homi Bhabha Road, Mumbai 400005, India\\}
\affilThree{\textsuperscript{3}University of Mumbai, Kalina, Mumbai-400098, India\\}
\affilFour{\textsuperscript{4}Dept. of Astronomy \& Astrophysics, Raman Research Institute,  Bengaluru-560080 India\\}
\affilFive{\textsuperscript{5}Inter-University Centre for Astronomy \& Astrophysics, Ganeshkhind, Pune-411007, India\\}
\affilSix{\textsuperscript{6}ISRO Inertial Systems Unit (IISU), Vattiyoorkavu, Trivandrum-695013, India\\}
\affilSeven{\textsuperscript{7}ISRO Satellite Center, HAL Airport Raod, Bengaluru-560017, India\\}



\twocolumn[{

\maketitle



\corres{jsyadav@tifr.res.in}

\begin{abstract}
Large Area X-ray Propositional Counter (LAXPC) instrument  on {\it AstroSat} is aimed at providing  high time resolution X-ray observations in 3-80 keV energy band with moderate energy resolution. To achieve large collecting area, a cluster of three co-aligned identical LAXPC detectors, is used to realize an effective area in access of $\sim$6000 cm${^2}$ at 15 keV. The large detection volume of the LAXPC detectors, filled with xenon gas at $\sim$2 atmosphere pressure, results in detection efficiency greater than 50\%, above 30  keV. In this article, we present salient features of the LAXPC detectors, their testing and characterization in the laboratory prior to launch and calibration in the orbit. Some preliminary results on timing and spectral characteristics of a few X-ray binaries and other type of sources, are briefly discussed to demonstrate that the LAXPC instrument is performing as planned in the orbit.
\end{abstract}

\keywords{space vehicles: instruments --- instrumentation --- X-ray --- detectors}

}]

\section{Introduction}    
{\it AstroSat} is India$^,$s first satellite fully devoted to astronomical studies. It is a multiwavelength observatory which covers 5 decades in energy from $\sim$1eV to 100 keV by means of 4 coaligned instruments capable of simultaneous observations in optical, Near-UV (NUV), Far-UV (FUV), soft X-ray (0.3--8 keV) and hard X-ray (3--100 keV bands). There is also a Scanning Sky X-ray Monitor (SSM) to monitor variability of known and new cosmic X-ray sources in 2--10 keV. An early description of {\it AstroSat} instruments and their characteristics can be found in Agrawal (2006) and a more recent account has been given by Singh et al. (2014). Multiwavelength observation capability is a unique feature of {\it AstroSat} which distinguishes it from other space observatories. Ability to study a cosmic source simultaneously in several spectral regions can probe origin of radiation in different spectral regions, their emission regions and processes and interrelationship among them. This is especially valuable in the studies of Active Galactic Nuclei and black hole X-ray binaries, which radiate over a broad spectral region and have complex multi-component energy spectra.

In this article  we focus on the LAXPC instrument, its salient characteristics, performance in orbit and highlight some of the results obtained during the performance verification phase. 

\section{Science Goals of LAXPC Instrument}
LAXPC instrument was conceived with the following objectives in mind :

\noindent $\bullet$ Intensity  variations over a wide range of time scales is the most prominent characteristics of accreting X-ray  binaries.  Neutron  binaries exhibit X-ray pulsations with periods of  $\sim$millisec to about 1000 sec. Detection and detailed  measurements of pulsations and binary periods, study of pulse profiles and their energy dependence, measurement of spin--up and spin--down rates of the pulsar and their time variations are important to probe details of the accretion process, geometry of the system and accretion torque acting on the neutron star.\\

\noindent $\bullet$ Quasi periodic Oscillations (QPOs) are a common feature of most black hole and neutron star Low Mass X-ray Binaries (LMXBs). They have so far been studied mostly below 10 keV due to lack of sensitivity at higher energy. LAXPC with its large collecting area at higher energy and time tagging of each event to 10 $\mu$sec accuracy, will be able to detect low and high frequency QPOs (up to a few kHz) in LMXBs and study time lag between QPOs of different energies  (Yadav et al. 2016b). Detection of high frequency QPOs in black hole binaries will provide a tool to probe spins of the black holes.\\

\noindent $\bullet$ Due to its wide spectral response (3--80 keV) and fair energy resolution ($\sim$ 12\% FWHM at 30 keV) this instrument will measure multi-component continuum spectra of all types of X-ray sources and detect break energy in X-ray pulsars. An important goal is to detect Cyclotron Resonant Scattering Features (CRSF) in the spectra of X-ray pulsars, which typically occur in 10 to 60 keV region, to  probe magnetic fields of the accreting neutron stars. Detailed study of variation in the energy of  CRSFs with the X-ray luminosity, pulse phase dependent changes in the line profiles etc are within capability of LAXPC.\\

\noindent $\bullet$ Binary periods and their evolution with time, measuring spectral evolution during Type I and Type II  X-ray bursts, study of thermonuclear bursts and burst oscillations in the rising and decay phase are some additional objectives.

 Realization of these objectives requires an instrument with wider spectral response having low energy threshold of 2--3 keV, large collecting area at low and high energies ($>$ 15 keV), and capability to measure arrival time of each photon accurate to about 10 $\mu$sec or better. The highly successful Proportional Counter Array (PCA) on Rossi X-ray Timing Explorer (RXTE) had some of these features but its spectral response was essentially limited due to rapid fall in the effective area of the PCA above 15 keV. The Imager on-Board INTEGRAL Satellite (IBIS), a Cadmium Telluride array based instrument on INTEGRAL and Burst Alert Telescope (BAT) on Swift have large effective area in hard X-ray band (20--200 keV) but their low energy threshold is about 15 keV and timing accuracy is also limited.  Due to this they miss detecting most of the photon flux that lies below 15 keV severely constraining their capability for measuring rapid variability like high frequency QPOs, millisecond pulsations etc. LAXPC was designed to have all these features missing in the earlier missions.

\section{LAXPC Instrument and its Principal Features: }
Yadav et al. (2016a) have presented details of LAXPC instrument and its characteristics  (also see Yadav et al. 2017). A detailed account of  the detector  design and characteristics, will be discussed by Agrawal et al. 2017. Complete details of calibration of the three LAXPC units, comparison of the calibration results with those obtained by GEANT4 simulations, derivation of important parameters, background rates and their variations with the orbital parameters etc have been comprehensively discussed by Antia et al. (2017).

The LAXPC instrument consists of three identical proportional counters having multilayer geometry. Each detector has independent front-end and event processing electronics units as well as command controllable High Voltage (HV) unit to make three independent detector units to ensure redundancy. The LAXPC units have the following  characteristics summarized in Table 1.

\noindent $\bullet$ The energy range of LAXPC  covers  3--80 keV. The lower energy threshold arises from the cut--off due to negligible X-ray transmission of 50 $\mu$m thick aluminised Mylar  below 3 keV. The higher energy threshold arises from the detection efficiency which drops rapidly at higher energies. If one takes into account the escape of Xenon K--fluorescent photons, a substantial fraction of 80 keV events arise by interaction of $\sim$110 keV incident X-rays.\\
 \begin{figure} [tp]
\centering
\includegraphics[width=0.95\columnwidth,angle=0.0]{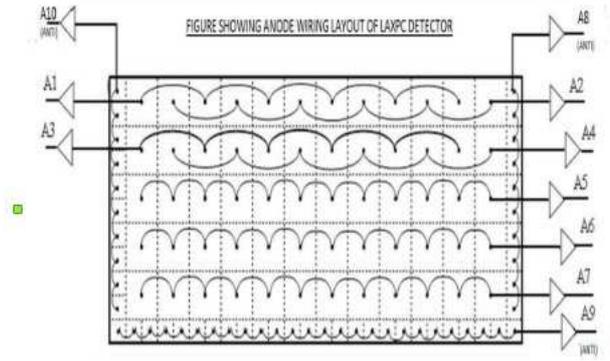}
\caption{LAXPC anode configuration, mutual anticoincidence and veto scheme.}
\label{fig1}
\end{figure}
\noindent $\bullet$  It  has high counting rate capability to study kHz scale intensity variations. This requires large photon collection area and high detection efficiency over the entire energy range.  Depending on the detection efficiency and transmission of the collimators, the effective area of the three units is $\sim$ 6000 cm$^{2}$ at  $\sim$ 15 keV and $\sim$ 5000 cm$^{2}$ at 50 keV (Yadav et al., 2016b,  Antia et al. 2017). The high source counting rate capability at 30 keV opens up the possibility of investigating properties of kHz QPOs in black hole binaries at energy  $>$ 20 keV and their time lags vis-a-vis lower energy photons.\\

\noindent $\bullet$  Xenon-Methane gas mixture at a pressure of 1520 torr  with detection depth of 15 cm  provides high detection efficiency up to 80 keV.\\

\noindent $\bullet$ Gas gain and energy resolution of the  detectors should not degrade in the orbit. Laboratory calibration shows typical energy resolution of the  detectors is about 12\% between 20 to 60 keV. Each LAXPC unit has an integrated on board Gas Purifier unit consisting of a DC motor driven bellow pump that circulates the detector gas, on activation of the pump by command, through an Oxisorb gas purifier to remove O$_{2}$, H$_{2}$O, CO$_{2}$ molecules and restore the energy resolution of the detectors. The gas gain and energy resolution of the detectors are constantly monitored by shining 59.6 X-rays from a highly-collimated Americium-241 source, only on one of the end Veto layer. 

 \begin{figure}
\centering
\includegraphics[width=0.95\columnwidth,angle=0.0]{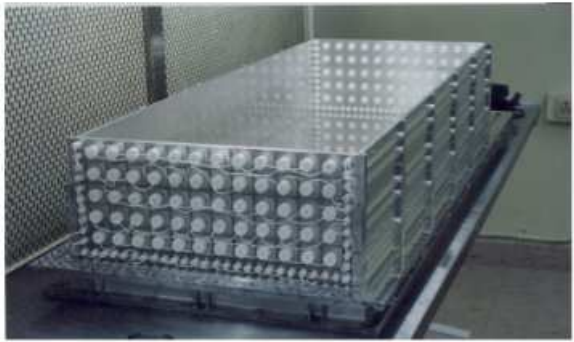}
\caption{A photograph of fully wired anode frame assembly of a LAXPC ready for installation in the detector housing.}
\label{fig2}
\end{figure}

\noindent $\bullet$ Cosmic ray interactions and Compton scatterings of diffuse cosmic X-rays and gamma-rays, leaking from all the sides of the detectors and  depositing energy in 3-80 keV, mainly contribute, to the non-cosmic X-ray background. To minimise this background the 60 X-ray detecting anode cells are arranged in 5 layers, as shown in figure \ref{fig1}. The ‘odd’ and ‘even’ anode cells in the top two layers are linked together to provide two outputs from each. In the remaining 3 anode layers, all the 12 anode cells of each layer are connected to bring one output from each layer as illustrated in the figure. Thus there are 7 signal outputs from the X-ray detection cells of each detector and they are all operated in mutual anticoincidence to reject events that produce signals in more than one anode simultaneously except when one of the two event is due to detection of Xe-K-fluorescent photon of 30 keV. Additionally, a Veto layer of 46 anode cells surrounds the X-ray detection volume on 3 sides, shown in figure \ref{fig1}, to reject interactions of cosmic rays and gamma-rays in the 3 side walls and anode frames. This scheme ensures that most of the non- X-ray produced events are rejected. A photograph of a fully wired and integrated Anode Frames ready for installation in the detector housing is shown in figure \ref{fig2}.

 \begin{figure}[bp]
\centering
\includegraphics[width=0.95\columnwidth,angle=0.0]{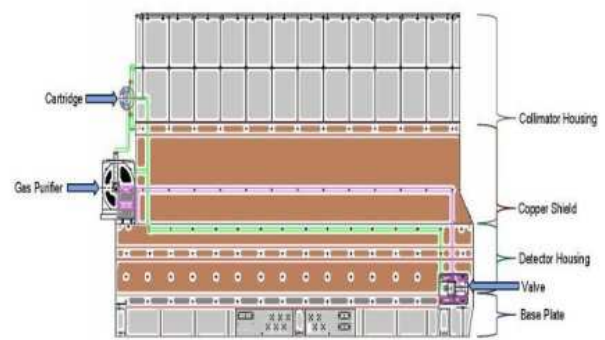}
\caption{Schematic diagram of LAXPC detector with major detector parts.}
\label{fig3}
\end{figure}

\noindent $\bullet$ The Collimator assembly of LAXPC consists of two parts (i) a Window Support Collimator (WSC) made from O.25 mm thick Aluminium slats of 7.5 cm  depth, with laser cut slits,. This is assembled to make square shaped honeycomb structure that supports the 50 $\mu$m thick Mylar film,  is housed in lower part of Collimator Housing as shown in a schematic view of LAXPC detector assembly in figure \ref{fig3}. The WSC has Field of View (FOV) of 5$^{o}$ x 5$^{o}$. The main FOV Collimator (FOVC) is made from a multilayer of tin, copper and aluminum glued together to make a 5 layer sheet (50 $\mu$m Al + 12.5 $\mu$m Cu + 50 $\mu$m Sn + 12.5 $\mu$m Cu + 50 $\mu$m Al) of 175 micron thickness and 37.0 cm width. Laser cut slits are made with a pitch of 7.0 mm and collimator is assembled to a square cell honey comb geometry to provide FOV of 1$^{o}$ x  1$^{o}$.  Each square shape cell has open area of 6.5 mm x 6.5 mm. The FOVC sits on top of the WSC in the collimator housing and the two are aligned carefully. Combination of WSC and FOVC results in LAXPC FOV of  0.9$^{o}$ x 0.9$^{o}$ at energy \textless 20 keV. A detailed description of the detector housing, collimators and other mechanical structures is given in Roy et al. (2016). The FOVC provides collimation at 90\% level at 80 keV but effectively blocks almost fully X-rays of lower energy entering from outside the FOV.



 \begin{figure} [tp]
\centering
\includegraphics[width=0.5\textwidth,angle=0.0]{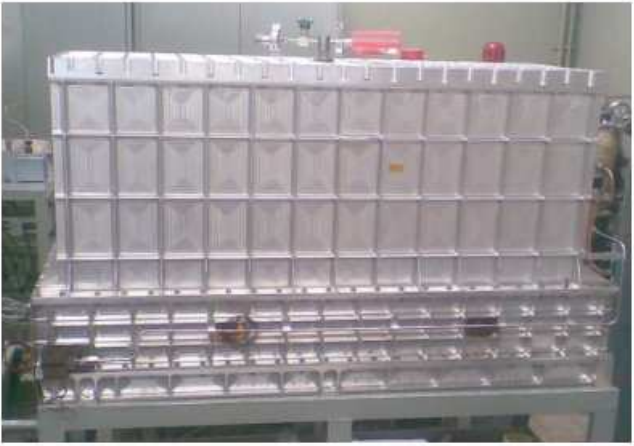}
\caption{Photograph of a  fully assembled LAXPC detector without tin shield.}
\label{fig4}
\end{figure}

\begin{figure}[th] 
\centering
\includegraphics[width=0.95\columnwidth,angle=0.0]{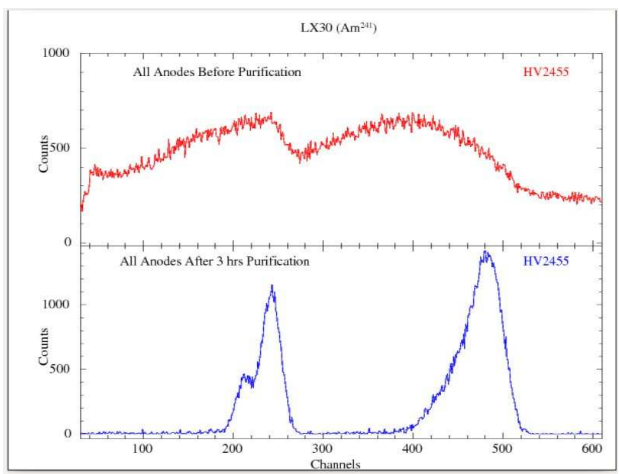}
\caption{Improvement in energy resolution of LAXPC30 after  purification for three hours in the lab before final calibration.}
\label{fig5}
\end{figure}

A complete description of collimator response to X-rays from cosmic X-ray sources based on scan across Crab Nebula, after the LAXPC instrument became operational in obit, has been provided by (Antia et al. (2017). A copper coated 1 mm tin shield is provided on the 5 sides of the LAXPC to prevent leakage of X-rays from the detector walls.   Photograph of a fully assembled LAXPC detector without tin shield is shown in figure \ref{fig4}.  The purification system was checked  in the lab before final calibration and improvement of detector energy resolution  is shown in figure \ref{fig5}  after purification of detector gas for 3 hours for LAXPC30 unit. 

\section{LAXPC Electronics :}
Each  detector has separate front-end electronics, a signal processing electronic unit, command controlled HV unit to control gas gain of the detector and a common time stamping electronic box called STBG to tag arrival time of each valid X-ray event. Only those single events that originate only from one of the five X-ray detecting anode layers (except those simultaneous signals that come from K-fluorescent X-ray) that have energy in 3-80 keV, not coincident with any veto signal, are processed for pulse height analysis. Such genuine events take 43 $\mu$sec to process before the electronics is ready to analyse another event (Yadav et al. 2016b). This is the dead time of each LAXPC. Pulse height of each valid X-ray event is analysed by a 1024 channel analyser and it is also time tagged. This information along with the layer of origin of the valid event, whether accompanied by a K-fluorescent event and various count rates viz veto layer rates, rate of below 3 keV events, 3--80 keV events, and above  80 keV events, rate of more than 2 simultaneous signals etc are recorded and telemetered to the ground. 

Each  detector has an independent HV DC-DC  Converter Unit whose output can be changed by command from zero to about 3000 V in coarse or fine steps of about 10 V.  The satellite passes through South Atlantic Anomaly (SAA) region which has a high concentration of protons and electrons. On  an average the satellite spends about 30 minutes in the SAA region. To prevent damage to the detector by high fluxes of the charged particles, the HV of the three  detectors is reduced by command to zero value when the satellite enters the SAA region based on the adopted SAA model (Antia et al. 2017). The HV is restored to normal value when the satellite exits the SAA zone. In each orbit during the pointing, when declination of the target source is \textless 60$^o$, the earth blocks the view axis direction for a duration that depents on the declination of the source. The usable source data per orbit  is available for only about 45 minutes on average.

 \begin{figure}[th]
\centering
\includegraphics[width=1.2\columnwidth,angle=0.0]{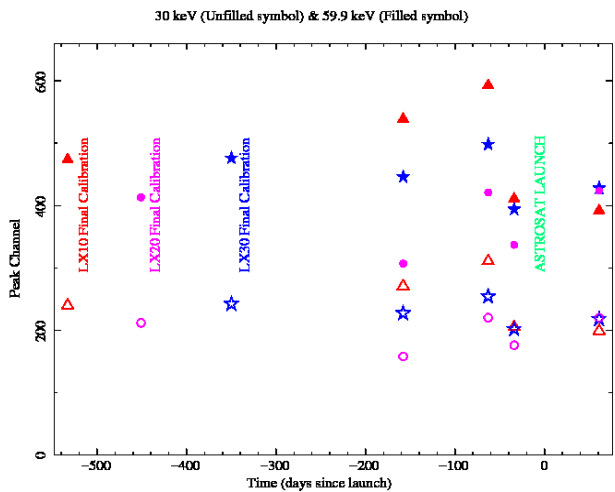}
\caption{Long term stability of three LAXPC detectors  tested for  almost two years. Position of 30 keV peak channel (unfilled symbols) and 59.6 keV peak channel (filled symbols) are plotted as a function of time since launch. }
\label{fig6}
\end{figure}

\section{Long Term Stability of LAXPC Detectors:}

All the LAXPC  units successfully completed space qualification tests during the year 2014. Each LAXPC detector was calibrated with monoenergetic X-rays from three radioactive sources in a thermovac chamber after the completion of the space qualification tests. We have studies long term stability and performance of the LAXPCs and the results are shown in figure 6. Final laboratory calibration of LAXPC10 unit was done in December 2013 and that of LAXPC30 was completed in October 2014. Here results for the three LAXPCs are shown for a period of about 2 years covering the time of final calibration in the laboratory to the first observation after activation and gas purification in the orbit on 28th November  2015. The date of final lab calibration is marked for three detectors in the figure. The 30 and 59.9 keV peaks are quite stable for LAXPC20 and LAXPC30 (within changes due to the impurity and purification of detector gas).  LAXPC10 shows very slow rise  in gain which is also observed in post {\it AstroSat} launch data (Antia et al. 2017).  This is most likely due to a very minor leak of gas from the detector. However, the leak rate is so small that the detector should function for the full life of 5 years with periodic adjustment of HV. LAXPC 30 has developed a significant leak in January 2016 due to which HV is corrected every week to restore the gas gain. This detector pressure may drop below 500 torr by the end of 2017 (Antia et al. 2017).

 \begin{figure}
\centering
\includegraphics[width=0.9\columnwidth,angle=0.0]{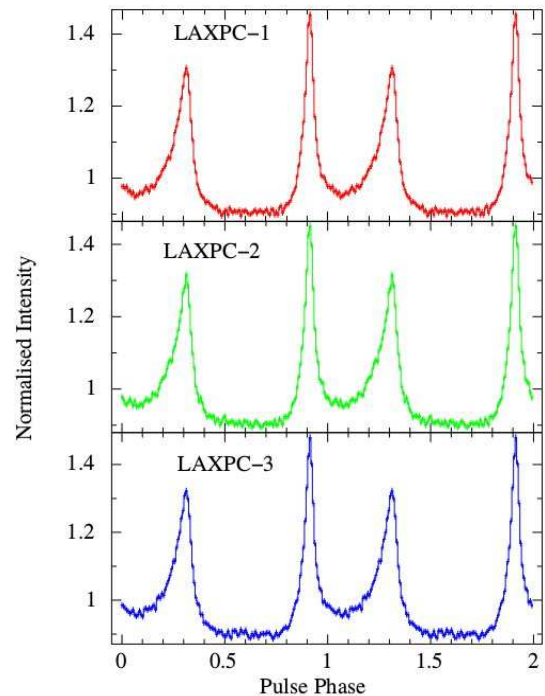}
\caption{The X-ray pulse profile of Crab Pulsar observed on 19$^{th}$ October 2015 (1$^{st}$  day of observation) in the three LAXPC detectors.}
\label{fig7}
\end{figure}

\section{Performance of the  LAXPC Instrument in Orbit:}
{\it AstroSat} was launched from Satish Dhawan Space Center in India on 28th September 2015 and placed in a 650 km circular orbit of 6$^{o}$ inclination to the equator.  A low inclination near equatorial orbit was chosen to achieve a stable background and minimize time spent in the SAA region. The low voltage of LAXPC processing electronic units and LAXPC detectors were activated on September 30 and October 1, 2015 respectively but the HV for the LAXPCs were turned on only on October 19, 2015 when the LAXPC instrument became fully operational. {\it AstroSat} was first pointed at the well known supernova remnant Crab Nebula; a bright hard X-ray source in the sky often used as a standard source for instrument calibration. The X-ray pulse profile of 33 millisec pulsations from the Crab pulsar obtained from observations on the very first day, is shown in figure \ref{fig7} for the three LAXPC detectors. All the known features of the Crab X-ray pulses viz main pulse and interpulse, their phases etc are seen in the figure validating the performance of the LAXPC instrument. 

\begin{figure}[t]
\centering
\includegraphics[width=0.95\columnwidth,angle=0.0]{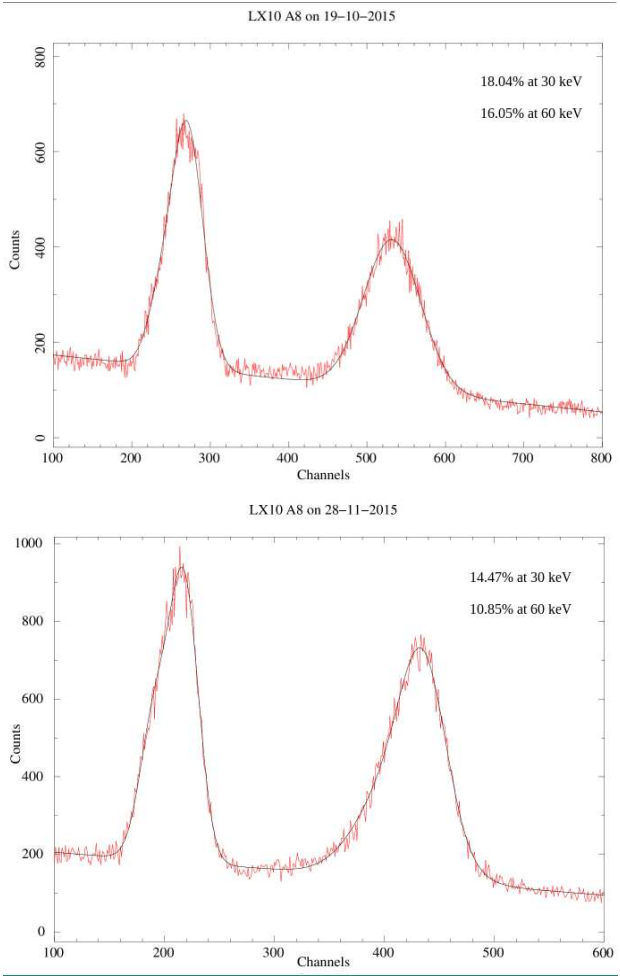}
\caption{Energy resolution of LAXPC10 before and after purification in the orbit.}
\label{fig8}
\end{figure}

Initial data showed that energy resolution of LAXPC10 detector had degraded due to prolong exposure to the atmosphere before launch caused by removal of the top cover several times for tests and after integration of the LAXPC instrument with the satellite bus. The top cover normally keeps the collimator volume under vacuum preventing exposure of the Mylar window to the ambient atmosphere. Therefore, gas purifier unit of each LAXPC was activated sequentially during October  and November 2015 to remove contaminants and restore the resolution of each detector. This is illustrated in figure \ref{fig8} where resolution of LAXPC10 is shown before and after purification of the gas. The LAXPC10 energy resolution improved from about 16\% to  about 11 \% at 60 keV.  At 6 keV,  energy resolution of 20\% is achieved as measured  from  Iron line (6.4 keV) seen in Cas-A  supernova Remnant (Yadav et al. 2016a).

 \begin{figure}
\centering
\includegraphics[width=0.5\textwidth,angle=0.0]{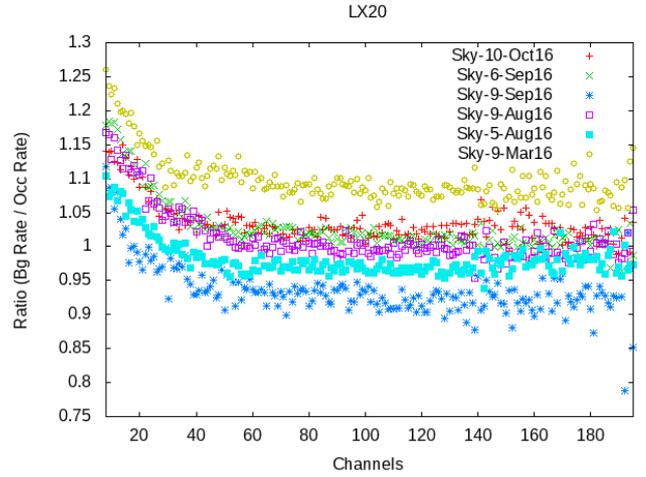}
\caption{ Ratio of background to the earth occultation counts for background observations 
during March to October 2016 for LAXPC20 detector.}
\label{fig9}
\end{figure}

After the activation of the LAXPCs, the background rates of the detectors were measured by pointing at the blank sky positions.  The Geant 4 simulations of the LAXPC detector show that dominant contribution to the background arises from Compton scattering of higher energy ($>$ 100 keV) diffuse gamma- ray background in and around the detector as well as by interaction of cosmic rays in the material around the detector. Details of this have been described by Antia et al. (2017). During the earth occultation the count rates drop to levels similar to that of the background but may include a small contribution of the earth X-ray albedo. To obtain the correct detector background rates, blank sky background observations were made during March-October 2016. The ratio of blank sky background and the earth occult counts (in the same orbit) is shown in figure \ref{fig9}. The data is binned in 256 channels. There is a significant difference in the count rates below 40 channels (\textless 15 keV) and hence the earth occultation counts can not be used as background for faint sources. Figure \ref{fig9} also shows temporal variations (see for blank sky 9 position) as well as significant differences in the count rates for different blank sky positions.  Development of a background model that can be used to compute the background rates from the latitude and longitudes of the satellite during source observation or alternately from the ULD (Upper level Discriminator set at 80 keV) count rates, has been described in detail by Antia et al. (2017).

\begin{figure}[t]
\centering
\includegraphics[width=0.5\textwidth,angle=0.0]{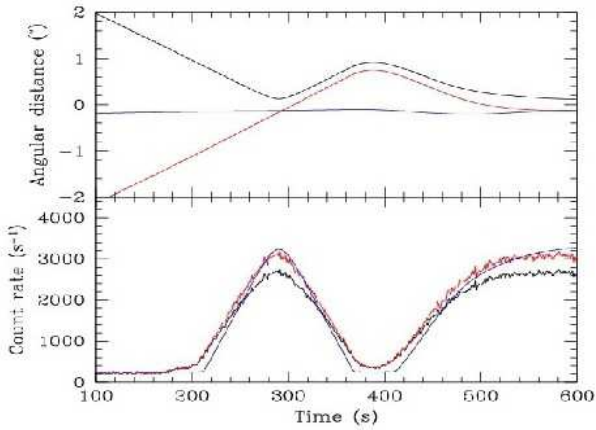}
\caption{Top panel shows plot of RA (red), DEC (blue) and net angle (black). Bottom panel shows the count rate observed (black), count rate corrected for dead time (red) and count rate expected from simulation of FOVC and corrected for dead time (blue).}
\label{fig10}
\end{figure}

The FOVC of each LAXPCs was calibrated by scanning slowly across Crab Nebula from -3 degree to +3 degree along RA and DEC. Scan was done with a speed of 0.01$^o$ per second from -3 RA to +1 RA as shown in figure \ref{fig10}. Such scan was repeated and offset of three detectors were estimated.  Details of the offset angles of the 3 LAXPCs are presented in Antia et al. (2017).


\section{Highlight of Some Initial Results:}
 \begin{figure}[bh]
\centering
\includegraphics[width=0.6\columnwidth,angle=-90.0]{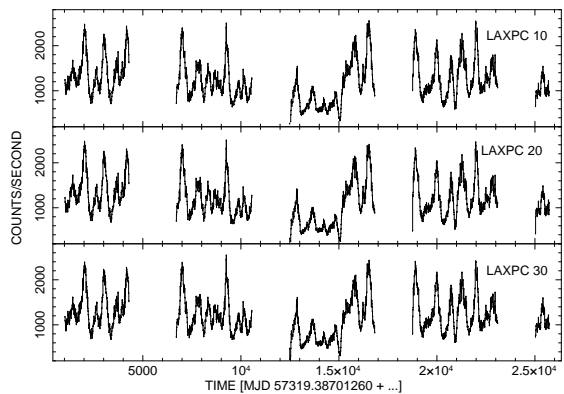}
\caption{Lightcurves of  4U 0115+63 in 3-80 keV band for three LAXPC detectors. Oscillating pattern is  clearly visible in all panels.}
\label{fig11}
\end{figure}
Soon after the full activation of the LAXPC instrument, there was an announcement (ATel 8279) about occurrence of a major Type II outburst in the Be X-ray binary pulsar 4U 0115+63 which has 24 day binary period and 3.62 sec pulsations. The {\it AstroSat} was, therefore, pointed at this source and LAXPC observations of this binary were made on October 24, 2015 when it was near its peak intensity (Yadav et al., 2016a).  Light curves of 4U 0115+63 for three detectors are shown (for 3-80 keV energy band) in figure 11.  It is apparent from the light curves that Quasi-periodic Oscillations (QPOs) with a period of  $\sim$1000 sec ( $\sim$1 mHz frequency) are present. Similar low frequency QPOs were also detected by the NuSTAR in 3-79 keV from this outburst one day prior and 4 days later to the LAXPC observations (Roy et al. 2017).  This is the lowest frequency QPO reported  so far from this binary.  A QPO of 1.46 mHz has been detected by Sidoli et al.(2015) in the High Mass X-ray Binary (HMXB) IGR J19140+0951 from XMM-Newton observation.  The intensity oscillations with period of $\sim$ 100 sec to 2.5 hour have been reported in the black hole nova V404 Cygni from optical and X-ray observations during its giant outburst in 2015 (Kimura et al.2016). 
Strong X-ray intensity oscillations  in GRS 1915+105 have been observed frequently; Taam et al. (1997) have reported first  such oscillations ($\rho$ class) with period of  $\sim$ 100 sec, Paul et al. (1998) and Yadav et al. (1999)  have reported such oscillations with period of 25 - 200 sec and Belloni et al. (1997) have reported  such strong  oscillations with period of  $\sim$ 1000 sec.  
Neilsen et al. (2012) have attempted to explain $\rho$ class oscillations  which are also referred as ``heartbeat'', in terms of a thermal–-viscous radiation pressure instability. Altamirano et al. (2011) has reported  detection 
of $\rho$ class in another black hole binary IGR J17091- 3624 (for comparison see Pahari et al. 2014).  The origin of these  oscillations is still not well understood specially  but likely to be due to  thermal-viscous instability in the inner accretion disk from which the matter accretes onto the neutron star or black hole. It was thought that the instability sets in when the accretion rate is close to the Eddington limit but detection of low frequency QPOs by Kimura et al. 2016 in V404 Cygni, when the accretion rate was much lower, is not consistent with this idea. Kimura et al. suggest the thermal instability in the inner accretion disk may be correlated to the long orbital period of the binary. A detailed discussion of the origin of $\sim$1 mHz QPOs in 4U 0115+63 has been given in Roy et al (2017).

\begin{figure}[th]
\centering
\includegraphics[width=0.7\columnwidth,angle=-90.0]{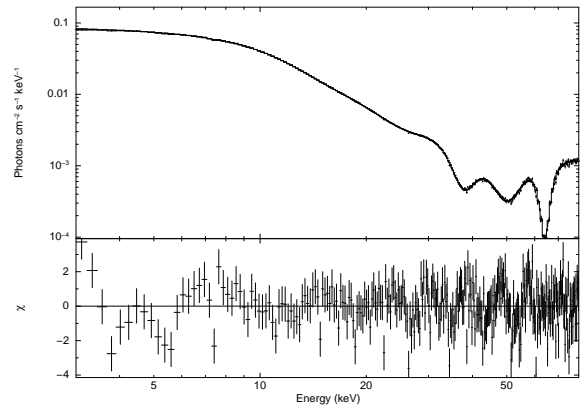}
\caption{3-80 keV energy spectrum  of 4U 0115+63 using laxpc10 data from observations on October 24, 2015.}
\label{fig12}
\end{figure}

CRSFs have been reported from 4U 0115+63 and from the measured value of fundamental line, magnetic field of the neutron star has been inferred. The energy spectrum of 4U 0115+63 was determined and spectrum obtained with LAXPC10 is shown in figure. \ref{fig12}. Presence of fundamental CRSF at $\sim$ 13 keV and its four higher harmonics can be clearly seen in the spectrum.
Model used for the fitting is Tbabs (gaussian + powerlaw) cyclabs *cyclabs *cyclabs *cyclabs *cyclabs. Powerlaw Photon Index is 0.367$^{+0.04}_{-0.03}$ with reduced chi square 1.4898 (272 degrees of freedom).  It may be pointed out that there is about 5\% uncertainty in the flux estimation due to systematics and background which may result in rather high chi square. The cyclotron line energy is estimated to be 13.47$^{+0.24}_{-0.30}$, 23.22$^{+0.82}_{-1.04}$, 37.31$^{+0.25}_{-0.29}$, 49.10$^{+0.56}_{-0.71}$, 63.99$^{+0.60}_{-0.53}$ keV (Roy et al 2017). Note that these are preliminary results and energies of fundamental CRSF and its higher harmonics may slightly change after refinement of energy calibration and response matrix.


Most X-ray binaries with a stellar-mass black-hole are X-ray transients in which matter is accreted from the companion star onto the black hole in a  sporadic way: they are characterized by a sudden and rapid rise in intensity (reaching a maximum in about a week) followed by a gradual decline. These outbursts typically last several months; eventually the source becomes too weak to be detected with current X-ray instruments. During an outburst, not only the intensity, but also the X-ray spectrum changes (Remillard \& McClintock 2006). In the bright phase of an outburst, the X-ray spectrum is dominated by thermal emission below $\sim$ 10 keV, most likely originating from an optically thick and geometrically thin accretion disk that extends down to the marginally stable orbit (Shakura \& Sunyaev 1973). The energy and timing characteristics change fast with evolution during the outburst. During  5-7 March, 2016, {\it AstroSat} observed  the black hole system GRS 1915+105  which was in the steep power law (SPL) state (Yadav et al. 2016b). In this state, X-ray flux does not change much at hour scale. In a  more recent classification of state, our observation will fall in hard intermediate state  (Belloni \& Motta (2016)). It may be pointed out that this system is not a typical black hole transient but rather a unique source which has been highly active since its first appearance. The event mode data, which give the arrival time and energy of each photon to a time-resolution of 10 microsec, are used to calculate the power density spectrum (PDS) for testing LAXPC timing characteristics up to the Nyquist frequency of 50 kHz.  The resulting PDS does not show any instrumental effect other than peaks beyond 10 kHz due to dead-time of the detector which is estimated to be 43 microsec (Yadav et al. 2016b).  Figure 13 shows PDS of GRS 1915+105 taken at an interval of 5 hours. Here each panel is for one orbit data which suggests  that LAXPC can study fast timing variability.


 \begin{figure}
\centering
\includegraphics[width=0.37\textwidth,angle=-90.0]{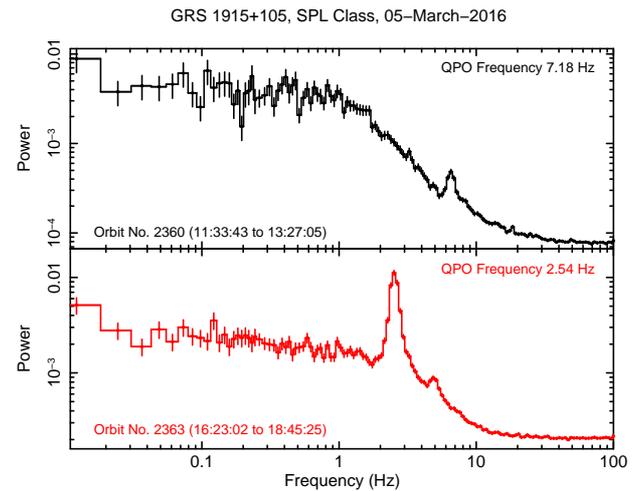}
\caption{Power density spectra  of GRS 1915+105  five hours apart.}
\label{fig13}
\end{figure}
\begin{figure}[bp]
\centering
\includegraphics[width=0.45\textwidth,angle=0.0]{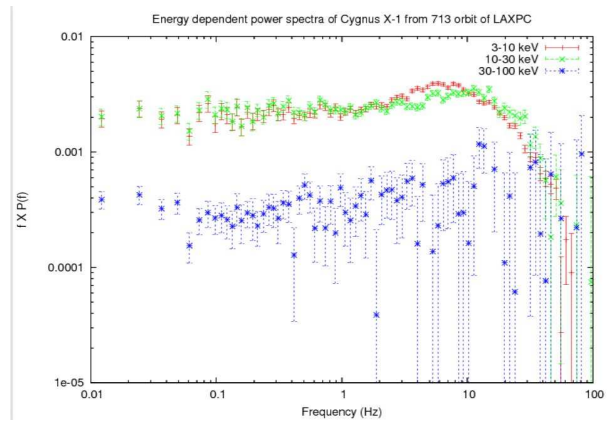}
\caption{Power density spectrum for Cyg X-1  in hard state  for single orbit LAXPC data in different energy bands.}
\label{fig14}
\end{figure}

During first week of January, 2016, {\it AstroSat} observed Cygnus X-1, the well known black hole X-ray binary in the low hard state which shows prominent thermal Comptonizaiton component.  The power spectrum can be characterized by two broad Lorentzian functions centered at $\sim$0.4 and $\sim$3 Hz  (for details  see Misra et al., 2017). Figure \ref{fig14} shows PDS power in energy bands 3-10, 10-30 and 30-80 keV.  The power in 30-80 keV energy band is  significant which  suggests that the  LAXPC instrument can cover larger energy range to study timing characteristics which was not possible with RXTE/PCA. 
  In conclusion it may be stated that LAXPC instrument has shown better capability to detect QPOs compared to earlier timing instruments and it can be used to study variation of QPO frequency as source evolves.  This improved sensitivity arises from larger effective area of the LAXPC instrument in hard X-rays as described by Antia et al. (2017) as well as from ability to record arrival time of each detected photon to an accuracy of 10 microsec without any onboard processing.


Some low mass X-ray binaries (LMXBs) in which the X-ray source is a neutron star with magnetic field \textless $10^9$ Gauss show thermonuclear bursts when accreting at a modest rate. X-ray temperature and flux variation during these bursts is the most commonly used method for measurement of radius of a neutron star. Such studies have so far been carried out in limited energy band and with assumption of simple black body type emission during the bursts. However, the non-burst emission from such neutron stars are known to suffer from reprocessing in the surrounding medium. With a wide energy band of the LAXPCs and large photon collection area, the thermonuclear burst spectroscopy will be investigated for signatures of reprocessing. LAXPC instrument has already observed  thermonuclear  bursts in two neutron star X-ray binaries; 4U 1728-34 and 4U 1636-536 (Chauhan et al. (2017), Paul et al. (2017)). During the  PV phase, the neutron star X-ray binary 4U 1728-34 was observed with LAXPC on 8th March 2016. We have detected typical Type-1 thermonuclear bursts in this source. Dynamical power spectrum of the data in the 3-20 keV band,  reveals presence of a high frequency QPO generally referred to as kilo Hz (kHz) QPO whose frequency drifted from $\sim$815 Hz at the beginning of the observation to $\sim$850 Hz just before the burst (Chauhan et al., 2017). The QPO is also detected in the 10-20 keV band, which was not detected in earlier {\it RXTE}/PCA observations of this source. 
 \begin{figure}[h]
\centering
\includegraphics[width=0.45\textwidth,angle=0.0]{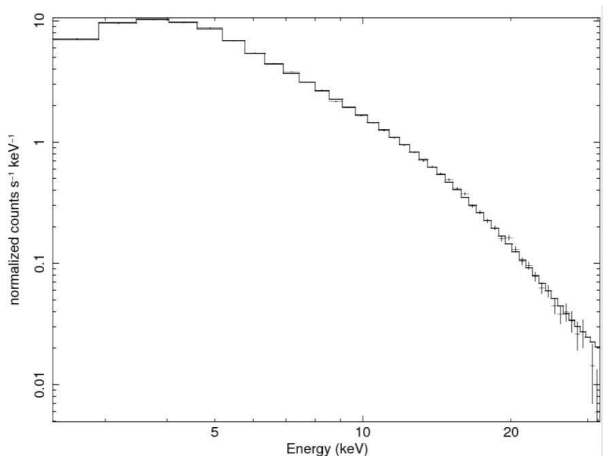}

\caption{Energy  spectrum  of Mkn421 a blazar  in 3-30 keV energy range observed in 
June 2016 with LAXPC20.}
\label{fig15}
\end{figure}

Several Blazar class AGNs have been observed with {\it AstroSat} so far. These are subclass of AGNs which include BL Lacs and FSRQs and have jets directed towards Earth within small angle around the line of sight (Urry \& Padovani 1995). Emission seen from Blazars is Doppler boosted. These are highly variable sources showing variability on time scales of minutes to years in various wavebands. Their Spectral Energy Distributions (SED) are characterised by double humped structure. First hump is attributed to Synchrotron emission from energetic electrons whereas origin of second hump is not clear. It could be leptonic or hadronic in origin (comprehensive review of these mechanisms can be found in (B\"{o}ttcher (2007)). X-ray observations play important role in understanding these enigmatic objects. We have preliminary results on some of the Blazars by now. One example is Mkn421 which is a nearby (z=0.031), bright and highly variable Blazar. LAXPC spectrum obtained from analysis of observations carried out in June 2016 is shown in figure 15. Spectrum over the energy range of 3-30 keV obtained from about 30 ks of data from top layer of LAXPC20 is shown in the figure. Spectrum is fitted with logparabola continuum along with line of sight absorption which is fixed to neutral hydrogen column density of 1.92 $\times$ 10$^{20}$ cm$^{-2}$. Logparabola parameters are $\alpha$ = 2.06$\pm$0.08 and $\beta$ = 0.33$\pm$0.04 which are consistent with the moderate flux level of the source (Sinha et al. 2015). Work on timing information for this source as well as similar work on other objects is underway.

\section*{Acknowledgment}
We acknowledge the strong support from Indian Space Research Organization (ISRO) in various aspects of instrument building, testing, software development and mission operation during payload verification phase. We acknowledge support of the scientific and technical staff of the LAXPC instrument team as well as staff of the TIFR Workshop in the development and testing of the LAXPC instrument.  We thank the referees for their comments and suggestions which improved the paper.

 
\begin{table*}
\centering
 \caption{Summary of the main characteristics of the LAXPC instrument}
\begin{center}
\scalebox{1.0}{%
\begin{tabular}{ |l|l| }
\hline
\hline
Number of LAXPC Units & 3\\
\hline
Size of X-ray detection volume & 100 cm long x 36 cm wide x 15 cm deep\\
\hline
Number of Anode cells & 60 arranged in 5 layers each with 12 Anode cells\\
\hline
Size of each Anode cell & 3 cm wide x 3 cm deep x 100 cm long\\
\hline
X-ray entrance window & 50 $\mu$m thick aluminized Mylar\\
\hline
Anode wire & 37 $\mu$m diameter gold coated stainless steel\\
\hline
Ground plane cathode wires & 50 $\mu$m diameter Beryllium-copper wires\\
\hline
Detector gas  & 90 \% Xenon and 10 \% Methane at total pressure of 1520 torr\\
\hline
Field of View & $\sim$ 0.9$^{o}$ x 0.9$^{o}$ at E $<$ 20 keV\\
\hline
Energy response & 3--80 keV\\
\hline
Typical energy resolution & $\sim$ 12\% FWHM at 22 keV (almost energy independent above 22 keV)\\
\hline
Effective area in 3--15 keV & $\sim$ 6000 cm$^{2}$\\
\hline
Effective area in 33--60 keV & $\sim$ 5000 cm$^{2}$\\
\hline
\hline
\end{tabular}}
\end{center}
\end{table*}

\end{document}